# Topological in-plane polarized piezo actuation for compact adaptive lenses with aspherical correction

F. Lemke, M. Stürmer, U. Wallrabe, M.C. Wapler
University of Freiburg, Department of Microsystems Engineering – IMTEK, Laboratory for Microactuators

**Abstract:**
In this contribution, we investigate the effects of using in-plane polarized piezo actuators with topological buckling displacement to drive glass-piezo composite membranes for adaptive lenses with aspherical control. We find that the effects on the focal power and aspherical tuning range are relatively small, whereas the tuning speed is improved significantly with a first resonance of 1 kHz for a 13 mm aperture lens.

**Introduction**

Adaptive lenses usually have a large outer diameter compared to their aperture and are limited to tens of ms response time (e.g. [1]). In [2, 3] we demonstrated piezo actuated ultra-compact lenses with aspherical correction and ms-scale response. These consist of an ultra-thin glass membrane sandwiched between two out-of-plane polarized ("$d_{31}$") piezo rings (Fig. 1a) that cover a flexible oil-filled fluid chamber made out of polyurethane as shown in Fig. 2 (with the new piezo design). The supporting ring in the fluid chamber acts both as a hinge for the membrane and as a soft spring to give way to the volume displacement.

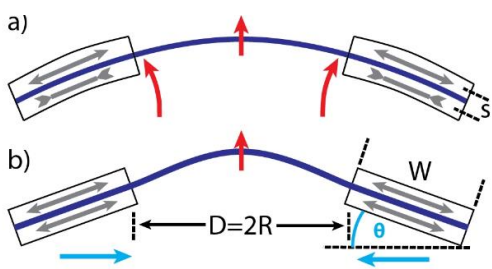

In our new approach we replace the $d_{31}$-piezo actuators by in-plane polarized films with ring-shaped finger electrodes, both with single-sided and with double-sided electrodes. This leads to the overall lens structure shown in figure 2. On the one hand, it allows for single-sided contacting that may simplify the fabrication. On the other hand, these actuators have an intrinsic buckling effect and potentially larger, anisotropic strains.
We will investigate whether the topological buckling effect [4] of the piezos does influence the deflection and aspherical control. We further investigate the difference between single-sided electrodes with easier fabrication and double-sided electrodes with potentially larger deflection [4, 5]. Furthermore, we analyse the resonance behaviour of the lens.

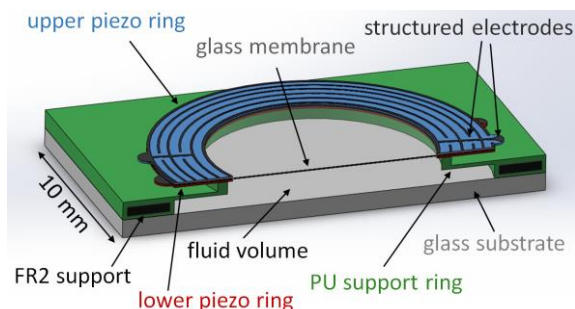

***Fig. 2:*** *Cross-section (schematic layout) of new lens with in-plane polarized piezo rings*

**Theory**

When applying an effective radial electric field E to the rings along the radial polarization, they contract tangentially and hence also overall radially with

$$\Delta R = R d_{31} E ,\qquad(1)$$

which is the same as for the conventional actuators. The width of the rings however expands with

$$\Delta W = W d_{33} E .\qquad(2)$$

By taking into account the finite width of the rings, we can apply the results of [4] that they will tilt conically with

$$\theta_{ring} = \pm\sqrt{2(d_{33} - d_{31})E} .\qquad(3)$$

While this result was derived for a closed disk, the mathematical locality of the derivation in [4] means that it applies also for an open ring.

In [2,3], we found that the active membrane has two actuation principles: The buckling and the bending mode. In the buckling mode, both piezo actuators contract with electric fields $E_{upper}$ and $E_{lower}$, such that the inner membrane buckles up- or downward

***Fig. 1:*** *Cross-section of glass (blue) and piezo rings (black) and their actuation principle: bending mode (a) and buckling mode (b)*

(Fig 1b). Assuming an infinitely thin membrane with aperture D=2R that displaces spherically, the curvature is

$$r^{-1} \approx \pm R^{-1}\sqrt{-3d_{31}(E_{\text{upper}} + E_{\text{lower}})}, \quad (4)$$

i.e. considering a refractive index n of the optical fluid we get the focal power

$$f^{-1} \approx \pm 2D^{-1}(n-1)\sqrt{-3d_{31}(E_{\text{upper}} + E_{\text{lower}})}. \quad (5)$$

In contrast to [2, 3], however, we now also have from eq. (3) the slope of the piezo rings

$$\theta_{ring} = \pm\sqrt{(d_{33} - d_{31})(E_{\text{upper}} + E_{\text{lower}})}, \quad (6)$$

which matches the edge of a spherical profile if $d_{31} = -\frac{1}{2}d_{33}$. In practice, most PZT materials are relatively close with $d_{31} \approx -0.4\, d_{33}$.

In the bending mode, we expect a curvature of

$$r^{-1} \approx \pm s^{-1} d_{33}(E_{\text{upper}} - E_{\text{lower}}), [2] \quad (7)$$

by simply taking into consideration the change of the width of the piezo rings, where *s* is of the order of the piezo and membrane thickness. Again, we can derive the focal power for this actuation mode

$$f^{-1} \approx s^{-1}(n-1)\, d_{33}(E_{\text{upper}} - E_{\text{lower}}). \quad (8)$$

For our dimensions, the buckling mode will dominate in a mixed mode and is mainly responsible for the deflection whereas the bending mode primarily affects the aspherical behavior with $(E_{\text{upper}} - E_{\text{lower}})$ as a control parameter.

In comparison to the old design [2] we expect a similar displacement or slight increase (due to the additional tilt of the rings) for the buckling mode, whereas the $d_{33}$ instead of $d_{31}$ in equation (7) leads to a lager displacement in the bending mode.

Unfortunately the inhomogeneous electric field generated by the surface electrodes again weakens this effect by approx. 30 % as shown in [5].

**Fabrication**

The piezo sheets (thickness $t = 100\,\mu m$, $d_{31} \approx -270 \times 10^{-12}\,m/V$, $E_C \approx 1\,kV/mm$) with silver electrodes are harvested from commercial sound buzzers. We cut them and structure the electrodes (400 µm distance, 100 µm width, Fig. 3) using a UV laser (355 nm, 2 W optical power); with a subsequent HNO$_3$ dip to remove residues that are caused by the laser ablation.

These piezo rings are heated above the Curie-temperature to T= 420°C ($T_c \approx 250°C$) to depolarize them. In the next step we glue a laser-structured glass membrane (Schott D263T, 50µm thickness) manually in between the piezo rings using hard polyurethane (Smooth-on Crystal Clear 200).

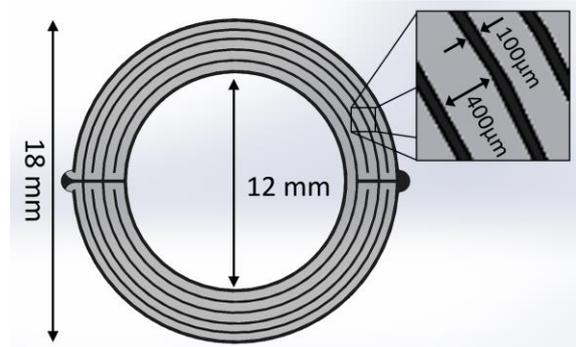

*Fig. 3:* *Electrode design: Piezo rings (grey) with ring shaped electrodes (black)*

For the mechanical support of the membrane, we cast the supporting ring and lens chamber from soft polyurethane (Smooth-on Clear Flex 50 with $E \approx 2.48$ MPa) with an enclosed FR2 stiffening structure and glue the piezo-glass-sandwich on top using the same soft polyurethane (Fig. 4). Finally, this lens chamber is filled upside-down with paraffin oil (refractive index n = 1.48, viscosity 20 mm²/s) and sealed by gluing a 0.5 mm thick glass substrate on top, again using the soft polyurethane. We add a custom-made PCB to contact the piezo rings to external wiring and also provide a mechanical support structure. To create the in-plane polarization we finally apply a high electric field of E≈1200V/mm for several minutes to the surface electrodes.

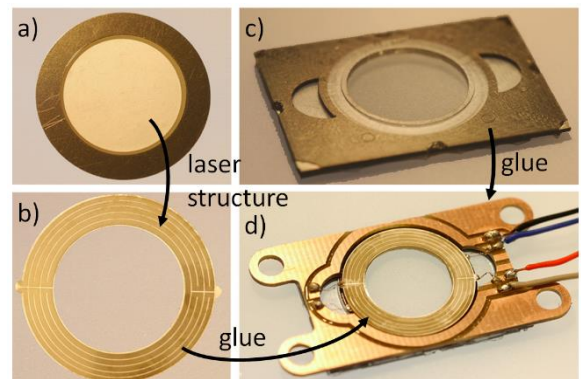

*Fig. 4:* *Some steps of the fabrication process: a) raw piezo on sound buzzer, b) structured piezo, c) soft support structure with FR2 stiffening, d) finished lens with PCB and electric contacts*

**Characterization**

To characterize the lenses, we scanned the surface with a confocal distance sensor and applied the lensmaker's formula to the measured profile. Assuming rotational symmetry, the lens surface can be approximated up to 4$^{th}$ order by

$$h(r) = \alpha_0 + \frac{f}{n-1}r^2 + \alpha_4 r^2, \qquad (9)$$

where $\alpha_0$ is a constant offset, $f$ is the focal power and $\alpha_4$ is a value for the aspherical behavior.

We used a quasi-static actuation (1 Hz) between $-\frac{E_C}{3}$ and $\sim 1.2 \frac{kV}{mm}$, corresponding to -40 V to 120 V for the old $d_{31}$ design and nominally to –160 V to 480 V (limited to –90 V up to 350 V by the amplifier) for the new design if we assume a perfectly homogeneous field between the electrodes. In the buckling mode both piezos were driven with the same voltage, i.e. $E_{\text{upper}} = E_{\text{lower}}$, whereas in the bending mode $E_{\text{upper}} = -E_{\text{lower}}$.

We measured three different lenses. A filled lens with the old d$_{31}$ design, a filled and an unfilled lens with single-sided electrodes in the new design and an unfilled lens with double-sided electrodes. For the unfilled lenses, we assumed n=1.48 to calculate a virtual focal power to be able to compare them with the filled lenses.

In Fig. 5, we see that the membranes show strong pre-deflections due to the remanent strain after polarizing, which makes it difficult to actually attain a flat state $r^{-1} = f^{-1} = 0$.

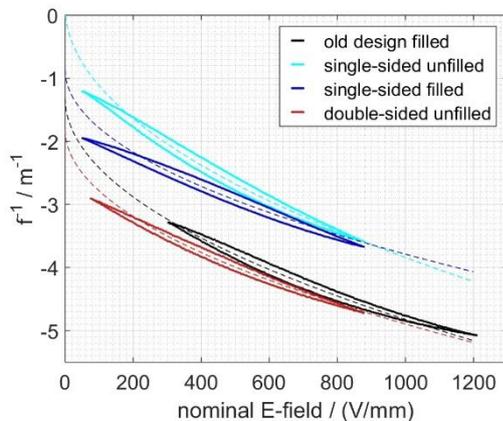

*Fig. 5:* *Focal power in the buckling mode with fitted curves according to the theory $f = a\sqrt{E} + b$ (dashed).*

As expected, the double-sided layout has a stronger displacement than the single-sided version (red and cyan). The filled single-sided lens has a reduced focal tuning range but also a higher pre-deflection compared to the unfilled single-sided membrane. Both effects result from the counter pressure of the fluid from the actuation and from the filling process.

Fig. 6 shows that the focal power range in the bending mode is much smaller than in the buckling mode and that it is more strongly reduced after filling (compare cyan and blue).

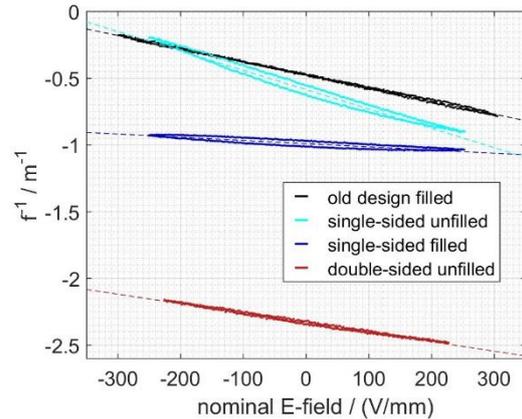

*Fig. 6:* *Focal power in the bending mode and linear fits.*

To determine the maximum working range (focal power and aspherical parameter), we actuated the lens in a trajectory that combines the most extreme buckling and bending modes. Depending on the actuation history and the back reaction forces, the lens membrane may reach a bi-stable point in the pure buckling mode and flip form positive focal power to negative focal power and vice versa. To take this into account, we applied the trajectory in both directions as shown in figure 7 (right graph).

The determined focal power and aspheric parameter are shown in the left graph of Fig. 7, where the enclosed area within the curves represents the overall working range of each lens. We find a different maximum focal power (extent left to right) as well as the associated minimum and maximum aspheric parameter (vertical extent at a fixed focal power).

We see that the aspherical tuning does not change qualitatively compared to the old system and decreases slightly in range, i.e. the additional tilt angle of the rings does not affect much the membrane profile. The strong pre-displacement of the double-sided layout however prevents it from snapping to the opposite direction (red) as the single-sided un-filled membrane does (cyan curve). In the filled lenses, the bistability is probably prevented by the initial pressure from the filling process.

Looking finally at the frequency spectrum in fig. 8, we see that the new layout is, however, much faster with a first resonance at 1 kHz (2.5 kHz unfilled), compared to ~0.5 kHz (over-damped) we find for the conventional design. This is probably caused by the higher pre-deflection, which suppresses the first resonance mode [2] and may also be an effect of the reduced electrode mass.

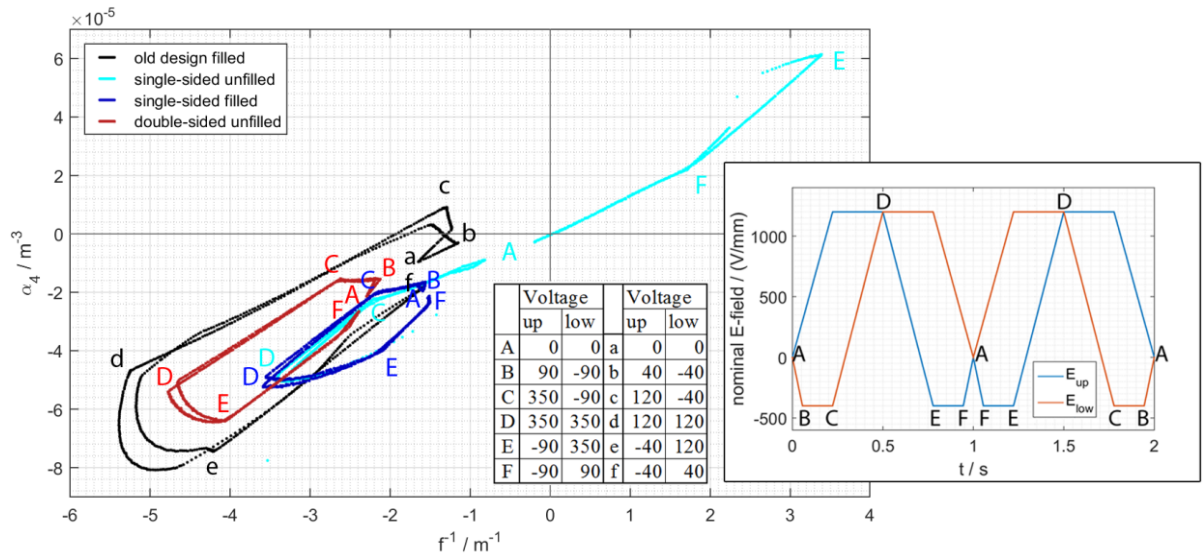

*Fig. 7:  Aspherical coefficient versus focal power. The trajectories A → B → C → D → E → F → A → F → E → D → C → B → A approximately outline the possible operation region with an applied electric field as shown in the right graph (which applies strictly only to the $d_{31}$ design).*

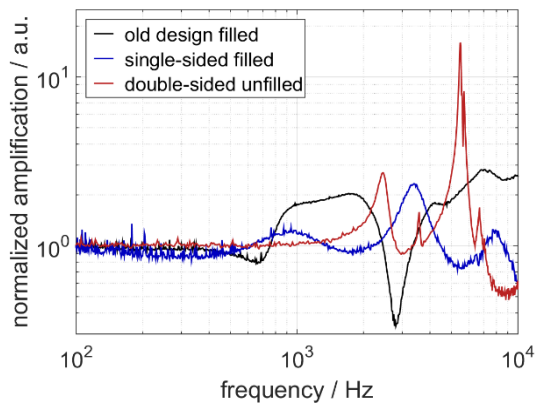

*Fig. 8:  Frequency spectrum*

## Conclusions

We successfully used the principle of in-plane polarized piezo actuation with a topological buckling effect to actuate adaptive lenses with aspherical correction. In a first attempt, with only one prototype per design, we have been able to show, that we can achieve a similar focal power and aspherical correction with the new concept like we had with the conventional design.

However, there was a small decrease in the aspherical tuning range as well as in the focal power range. It seems that the effects of the bending forces are stronger than the purely geometric theoretical improvement in the boundary angle of the membrane. Furthermore, we found in [5], that in the interdigitated electrode design, the effective strain is smaller than the one obtained from the piezoelectric coefficients and the nominal field strength, which can explain the decreased tuning range. Hence, the potential improvement over the conventional $d_{31}$ is in this case rather the simplification in the contacting if we use the single-sided design.

A precise quantitative comparison, however, would require better statistics and a more controlled fabrications process. We further expect that using a wider electrode spacing [5] could yield an improvement of ~30% for the focal power.

## Acknowledgments

We would like to thank Binal Bruno for supporting the confocal sensor measurements and the data processing of the obtained surface profiles.

This research is funded partly by DFG grants WA 1657/6-1 and the Cluster of Excellence BrainLinks-BrainTools EXC 1086.